\documentclass[12pt]{iopart}

\usepackage{iopams}  
\usepackage{aas_macros}
\usepackage{graphicx}
\usepackage{float}
\usepackage[toc,page]{appendix}
\usepackage{xfrac}
\usepackage{todonotes}
\usepackage{lipsum,graphicx}
\usepackage{cite}
\usepackage{multicol}
\usepackage{acronym}

\newacro{EOS}{equation of state}
\newacro{PN}{post-Newtonian}
\newacro{EOB}{Effective-one-body}
\newacro{BNS}{binary neutron star}
 
\begin{document}
\newcommand{\GWPAC}{\address{Gravitational Wave Physics and
  Astronomy Center, California State University Fullerton,
  Fullerton, California 92834, USA}}
\title[Matter Effects on Searches for Gravitational Waves]{Matter Effects on LIGO/Virgo Searches for Gravitational Waves from Merging Neutron Stars}
  \author{Torrey Cullen} \GWPAC \address{Louisiana State University
Baton Rouge, Louisiana 70803} \ead{ctorr23@lsu.edu}
  \author{Ian Harry} \address{Max Planck Institute for Gravitational Physics (Albert Einstein Institute), Am M{\"u}hlenberg 1, D-14476 Potsdam-Golm, Germany}\ead{ian.harry@ligo.org}
\author{Jocelyn Read} \GWPAC \ead{jread@exchange.fullerton.edu}
\author{Eric Flynn} \GWPAC \ead{eflynn67@csu.fullerton.edu}

\address{}
\begin{abstract}
Gravitational waves from merging neutron stars are expected to be observed in the next 5 years. We explore the potential impact of matter effects on gravitational waves from merging double neutron-star binaries. If neutron star binaries exist with chirp masses less than roughly $1$ solar mass and typical neutron-star radii are larger than roughly $14$\,km, or if neutron-star radii are larger than $15$-$16$\,km for the chirp masses of galactic neutron-star binaries, then matter will have a significant impact on the effectiveness of a point-particle-based search at Advanced LIGO design sensitivity (roughly 5\% additional loss of signals). In a configuration typical of LIGO's first observing run, extreme matter effects lead to up to 10\% potential loss in the most extreme cases.


\end{abstract}
\section{Introduction}

The first detections of gravitational waves from merging binary black holes \cite{2016PhRvL.116f1102A,2016PhRvL.116x1103A,PhysRevLett.118.221101} have ushered in the era of gravitational-wave astronomy.  
The Advanced Laser Interferometer Gravitational Wave Observatory (LIGO)\cite{TheLIGOScientific:2014jea}, shortly to be joined by advanced Virgo\cite{TheVirgo:2014hva} and KAGRA\cite{PhysRevD.88.043007}, is also sensitive to gravitational waves from systems with lower component masses, such as double neutron-star and black-hole/neutron-star mergers \cite{2016PhRvX...6d1015A, ratespaper}. While no mergers involving neutron stars have yet been detected,  upcoming observing runs with increasing sensitivity and duration will increase LIGO's chances of such an observation \cite{2016LRR....19....1A}. Astrophysical estimates of double-neutron-star merger rates, which are not yet constrained by the lack of LIGO detections, predict $\sim0.2$-$200$ detected mergers per year once LIGO reaches design sensitivity near the end of this decade \cite{2016LRR....19....1A,ratespaper}.

The detection of lower-mass \ac{BNS} mergers will require matched-filtering: comparing the detector output to a predicted signal \cite{2017arXiv170501845D} generated for particular source parameters. The first instance of a detected signal found through matched filtering was recorded in December of 2015 as a 22-solar-mass coalescence (GW151226) \cite{2016PhRvL.116x1103A}. LIGO's compact binary coalescence matched-filter searches cover binary component masses ranging from $[1.0,100]M_\odot$ for core searches \cite{2016PhRvX...6d1015A,2017arXiv170501845D}, and have also been used to search for higher-mass mergers \cite{2017arXiv170404628T}.

Early investigations of gravitational waves from merging neutron stars \cite{1992ApJ...398..234K,1992ApJ...400..175B,1996PhRvD..54.3958L} predicted that tidal interactions would be small until the final stages before merger, at high gravitational-wave frequencies. As a result, search strategies for gravitational waves from binary neutron stars were developed using point-particle inspiral models \cite{1993PhRvL..70.2984C,2012PhRvD..85l2006A}. The effectiveness of the point-particle search for initial LIGO was confirmed in Berti et. al. \cite{2002PhRvD..66f4013B}.  However, the impact of matter on the final cycles continued to drive interest in measuring neutron-star \ac{EOS} or radius from the detected signals \cite{2008PhRvD..77b1502F}. Matter effects are now studied with a combination of analytic 
and numerical 
investigations of the impact of the neutron-star equation of state on emitted gravitational waves. Their signature provides a possible measurement of equation-of-state related quantities. 
However, matter effects coming from the finite size of neutron stars are not included in the Advanced LIGO template banks---the set of waveform models that detected signals are compared to during matched-filtering.

In this work, we will explore the size of matter effects over the full low-mass region targeted by ground-based detector searches, component masses of $[1.0,2.0]M_\odot$, where both components are expected to be neutron stars and matter effects are most significant.

Neutron stars have been observed with well-measured masses ranging from 1.17 to 2.0\,$M_\odot$\cite{2016ARA&A..54..401O}, with those in double-neutron-star binaries bounded by the most unequal-mass system, J0453+1559, with component masses 1.174\,$M_\odot$ and 1.559\,$M_\odot$\cite{2015ApJ...812..143M}. With a narrow observed range of masses in mind, and due to computational limitations, many previous studies of the impact of matter effects have focused on canonical equal-mass systems with $1.35\,M_\odot$ or $1.4\,M_\odot$ components. 

In this paper, we explore the relative sizes of waveform effects to determine the potential impact of matter on Advanced LIGO searches and parameter estimation. We compare the impact of incorporating matter into the waveforms to that of underlying relativistic point-particle modeling choices, and we compare the impact of leading-order post-Newtonian matter effects to that of a numerically simulated neutron-star merger. We assess the importance of including matter effects for Advanced LIGO's \ac{BNS} search as well as for parameter estimation.

\section{Waveform models considered}
 
LIGO uses a number of waveform approximants: approximate predictions for the gravitational waves produced by an astrophysical system following the Einstein Field Equations. We focus on three choices in this study, named TaylorF2, TaylorT4, and EOBNRv2  in \textit{LIGO Algorithms Library} (LAL), which are available for LIGO/Virgo searches and analysis \cite{LALref}. 

\subsection{Waveforms without matter}

\begin{figure}[htb]
\hspace*{-.5cm} 
\includegraphics[width=7.0in]{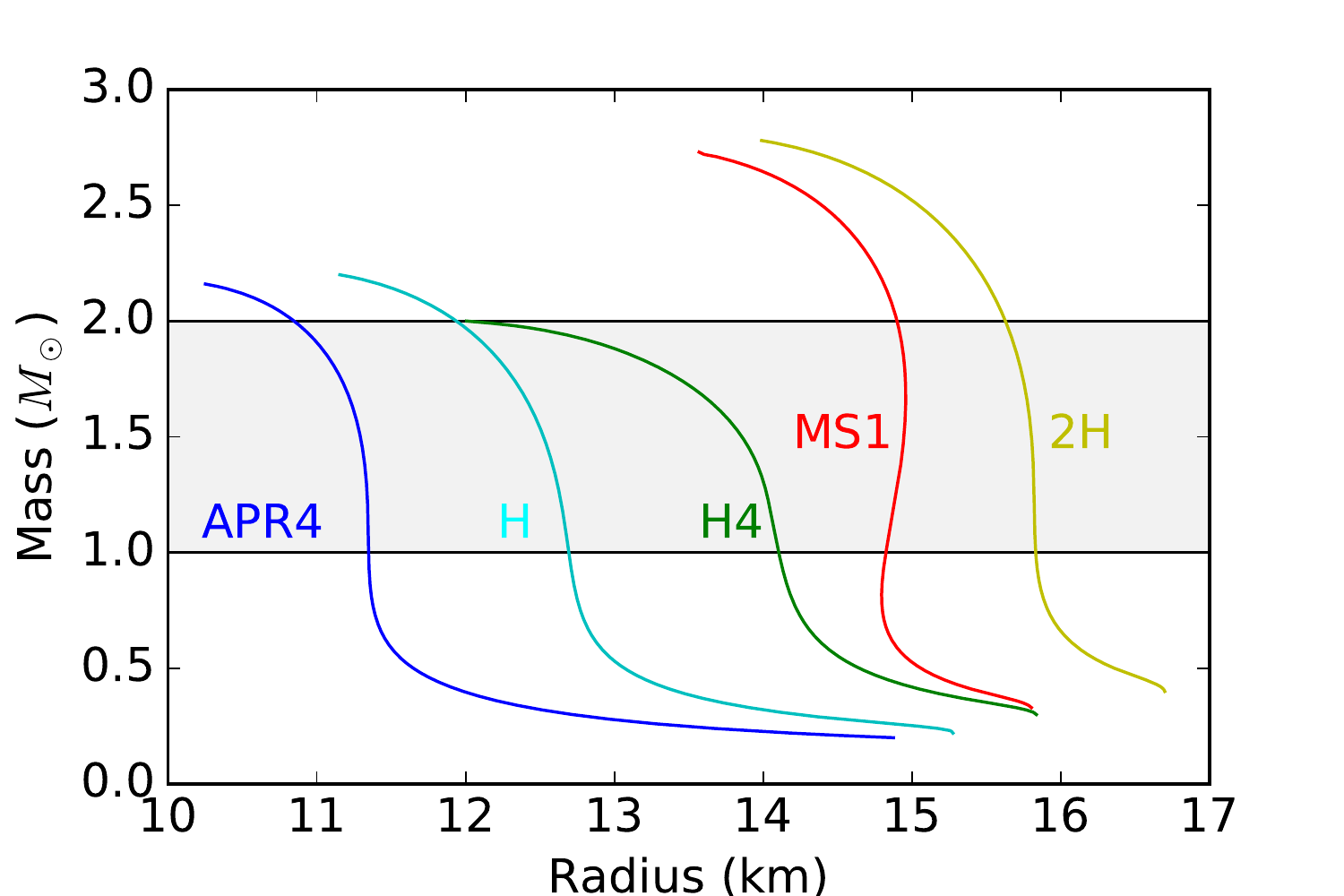}
\caption{Curves illustrating the mass-radius relationship of the stars for the given EOS. The gray band shows the full low-mass region that is explored throughout this paper. The mass-radius relationships are calculated using \textit{lalsim-ns-params} by specifying the four parameters that make up the piecewise polytrope: see Table 1}
\label{fig:massradius}
\end{figure}

TaylorF2 and TaylorT4 are, respectively, a frequency-domain and time-domain example of a \ac{PN} expansion,  used to quickly model the orbital phase of a system and the resulting gravitational waveform \cite{T4&F2}. By default, these approximants assume the components are point particles and that the objects are moving slow relative to the speed of light. These \ac{PN} expansions are calculated to specific orders, and because calculating higher order terms is difficult, different re-expansion techniques diverge when the absent higher order terms become important. Because gravitational wave signals from compact binary mergers chirp, or increase in frequency and amplitude as the system evolves, these approximants break down near merger as the velocity of the bodies approach the speed of light.
However, since the merger of neutron stars is at much higher frequency than LIGO's most sensitive region, the TaylorF2 approximant is the current model used to search for signals in the \ac{BNS} region of the parameter space ($m_1,m_2 \leq 2$). A second \ac{PN} choice, TaylorT4, gives a reference for the differences within \ac{PN} approximants when compared to TaylorF2. TaylorT4 was chosen in particular as it agrees well with numerical results when the mass ratio $q=\frac{m_1}{m_2} \approx 1$, which is true for the space we are looking at as $q_{max} = 2$ for \ac{BNS}. \cite{2016PhRvD..93d4064B}

Effective-One-Body (EOB)  models, or phenomenological waveforms calibrated to numerical simulation results, are required to accurately capture the underlying matter-free dynamics seen in a black hole merger. Here, we use an EOB approximant that is calibrated to numerical relativity over a wide parameter space as an accurate inspiral, merger, and ringdown for a matter-free binary black hole system \cite{EOB}.

\subsection{Waveforms with  matter}

We explore both the leading-order tidal contributions of matter on the gravitational waves and the additional impact of a fully relativistic merger from the numerical simulations of \cite{hotokezaka1}.

The TaylorT4 and TaylorF2 approximant have the option implemented for the \ac{EOS} to be set which allows tidally effected PN waveforms to be studied \cite{wadespaper}. The strength of the tidal contribution is specified by a dimensionless tidal deformability parameter, $\Lambda$. This tidal parameter is defined as

%
\begin{eqnarray}
\Lambda \equiv \frac{2}{3}k_2 \left(\frac{R}{M}\right)^5\label{eq:1}
\end{eqnarray}
where R is the radius, M is the mass, and $k_2$ is the quadrupole love number \cite{readspaper,2004PhRvD..69j4017P}. It is determined by the \ac{EOS} for a given mass of star. We determine the parameters for a span of equations of state that range from a moderate APR4 equation of state, compatible with many modern astrophysical constraints, as well as more extreme cases to show the effect of unexpectedly large-radii neutron stars on LIGO searches. Realistic EOS are specified in the piecewise polytrope formalism using three adiabatic indices and the pressure at the plane separating region 1 and 2 \cite{2009PhRvD..79l4032R}. The Equations of State used here are defined in \cite{readspaper,hotokezaka1,hotokezaka2}.

\begin{table}[htb]
\begin{center}
 \begin{tabular}{||c c c c c||}
 \hline
 EOS & $\log{P_1} (\frac{dyne}{cm^2})$ & $\Gamma_1$ & $\Gamma_2$ & $\Gamma_3$ \\ [0.5ex] 
 \hline\hline
 APR4 & 34.269 & 2.830 & 3.445 & 3.348 \\
  \hline
  H & 34.5036 & 3.0 & 3.0 & 3.0 \\
 \hline 
 H4 & 34.669 & 2.909 & 2.246 & 2.144 \\ 
 \hline
 MS1 & 34.858 & 3.224 & 3.033 & 1.325 \\
 \hline

 2H & 34.9036 & 3.0 & 3.0 & 3.0 \\
 \hline
\end{tabular}
\end{center}
\caption{Parameters of the piecewise polytrope EOS used in this work. These quantities can be specified in the \texttt{lalsim-ns-params} command to give the $\Lambda$ value of the star. \texttt{PyCBC}'s \texttt{get}\underline{ }\texttt{td}\underline{ }\texttt{waveform} accepts this $\Lambda$ quantity for certain PN approximants, allowing us to study the gravitational wave output of tidally affected stars.}
\end{table}

The mass-radius relationship of the EOSs considered in this paper are shown in Figure \ref{fig:massradius}.

Finally, PN approximants diverge near merger, and nonlinear interactions between the two neutron stars may produce additional effects on the orbits beyond the leading-order tidal contributions. Numerical simulations of Einstein's Field Equations with relativistic hydrodynamics capture the stars' full interaction as they merge, but are too expensive to generate or to make enough simulations to sufficiently cover the entire detectable parameter space. The current state of BNS numerical simulations organized by component mass and number of simulations is summarized in Figure \ref{fig:bns-sims}. Here, we include representative examples of full numerical merger in this study using hybrid waveforms constructed from the numerical mergers over a range of mass and mass ratio in \cite{hotokezaka1}. We employ the TaylorT4 post-Newtonian model for the early inspiral model, but rely for the last orbits and the post-merger on Numerical simulations.

\begin{figure}[htb]
\includegraphics[width=7.0in]{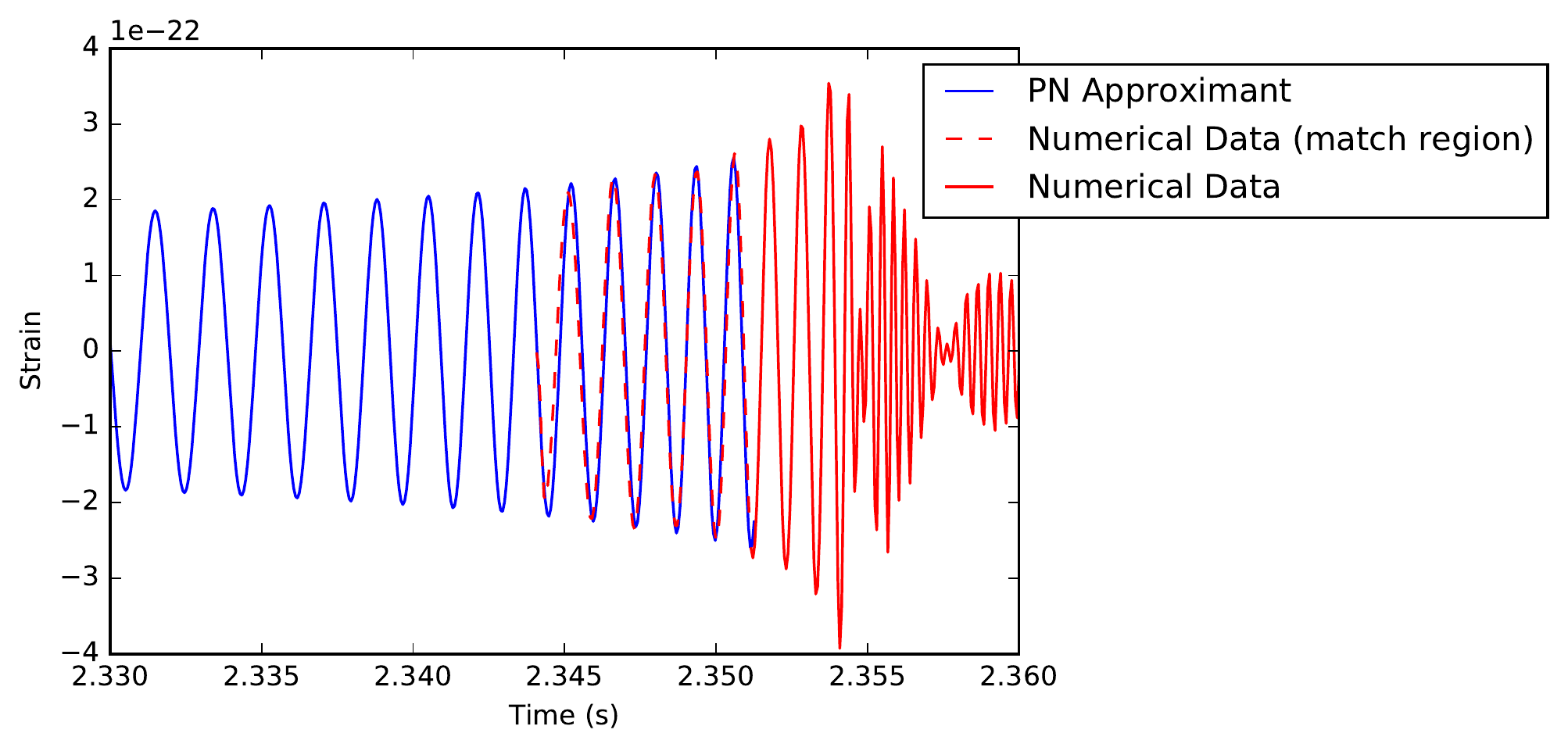}
\caption{Illustration of the construction of a hybrid waveform. The numerical simulation data is from \cite{hotokezaka1}. This is a zoomed in window; note that the PN part of the waveform extends to time t=0.}
\label{fig:hybrid}
\end{figure}

The hybrid waveforms used in this study were created by reading in an existing numerical simulation and `stitching' it onto the end of PN with the same parameters. The procedure roughly follows \cite{readspaper}, defining a match region where the waveforms are aligned to be maximally correlated, then linearly turning on the numerical waveform through a modified windowing function at the same rate the PN waveform is turned off,  over $\approx5-7$ cycles. An example hybrid waveform is shown in Figure \ref{fig:hybrid}. The values of each parameter for the corresponding EOS can be found in Table 1 \cite{hotokezaka2}.

\section{Size of waveform differences}
In order to make statements about which waveform modeling choices have the greatest effect on waveform variation and searches, we first define the ability to distinguish between two different waveforms. 

Take the noise weighted inner product between two waveforms $h_1$ and $h_2$, with a noise power spectral density $S_n (f)$,

\begin{eqnarray}\label{eq:prod}
\langle{h_1} | h_2 \rangle{}  \equiv 4 Re \int_{0}^{\infty} \frac{h_1^* (f) h_2 (f)}{S_n (f)} df .
\end{eqnarray} 
The characteristic signal to noise ratio is $\rho \equiv \langle{h} | h \rangle{}^\frac{1}{2}$. Two waveforms are estimated to be indistinguishable if the quantity \cite{2008PhRvD..78l4020L,readspaper},
\begin{eqnarray}
||\delta h|| \equiv ||h_2 - h_1|| \equiv \sqrt[]{\langle{h_2-h_1} | h_2 - h_1 \rangle{}} \leq 1.
\end{eqnarray} 
This quantity is essentially the signal to noise ratio of the difference between the two waveforms that are being compared. If this quantity is less than 1, the waveforms' differences are smaller than the noise. We record the maximum distinguishable distance as the distance at which $||\delta h|| = 1$, although of course it is not guaranteed that the effect will be distinguishable in practice at that distance. For comparison, we also record the maximum \emph{detectable} distance of a set of \ac{BNS} systems, where the characteristic signal to noise ratio is $\rho = 8$, also referred to as the horizon distance $D_{horizon}$.

\begin{figure}[htb]
\hspace*{-.75cm} 
\includegraphics[width=6.5in]{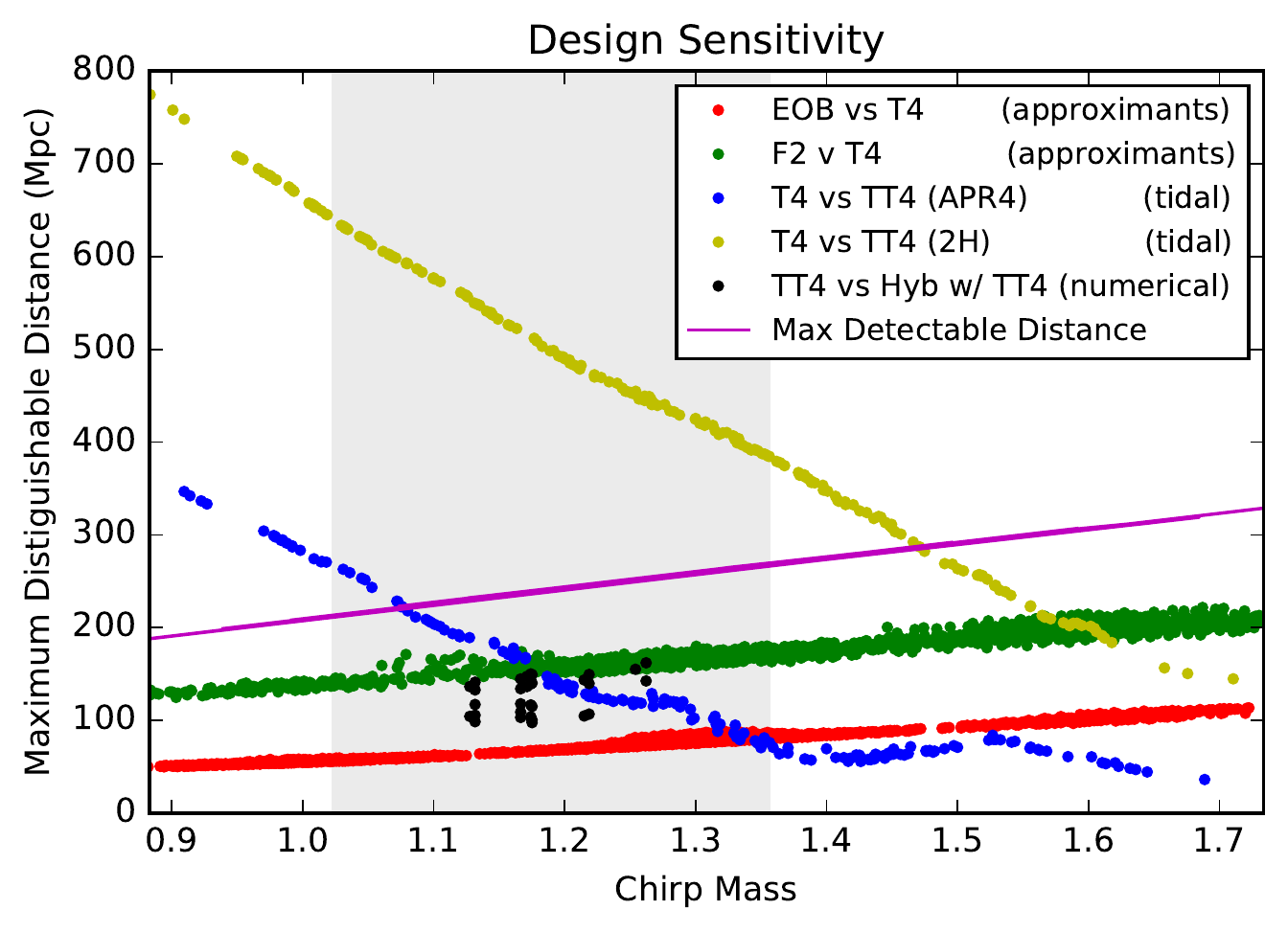}
\caption{Maximum distinguishable distance between two given system as a function of chirp mass. The purple solid line is the max detectable distance for a BNS system at design sensitivity corresponding to a signal to noise ratio of 8. The gray region represents the chirp mass of of previously observed BNS systems \cite{2016ARA&A..54..401O}. Note that a 1.4,1.4 $M_\odot$ system corresponds to a chirp mass of $\approx 1.22$. Waveform models used in this figure include the Effective One Body (EOB), TaylorF2 (F2), TaylorT4 (T4), and a tidal TaylorT4 (TT4).}
\label{fig:disting}
\end{figure}

In order to isolate which effects contribute the most to waveform variation, we calculate these distances for a large set of example binary neutron star systems where component masses are uniformly drawn from the mass range  $[1.0,2.0]M_\odot$. Component masses are held constant between $h_1$ and $h_2$ and component stars are not given spin. In Figure $\ref{fig:disting}$, characteristic distances are plotted as a function of the chirp mass of the system, where chirp mass is given as

\begin{eqnarray}
\mathcal{M}_c = \frac{(m_1m_2)^\frac{3}{5}}{(m_1+m_2)^\frac{1}{5}}.
\end{eqnarray}
\ref{sec:chirpvstotal} compares results in terms of total mass and mass ratio. 
%
%

\begin{figure*}[htb]
\includegraphics[width=7.0in]{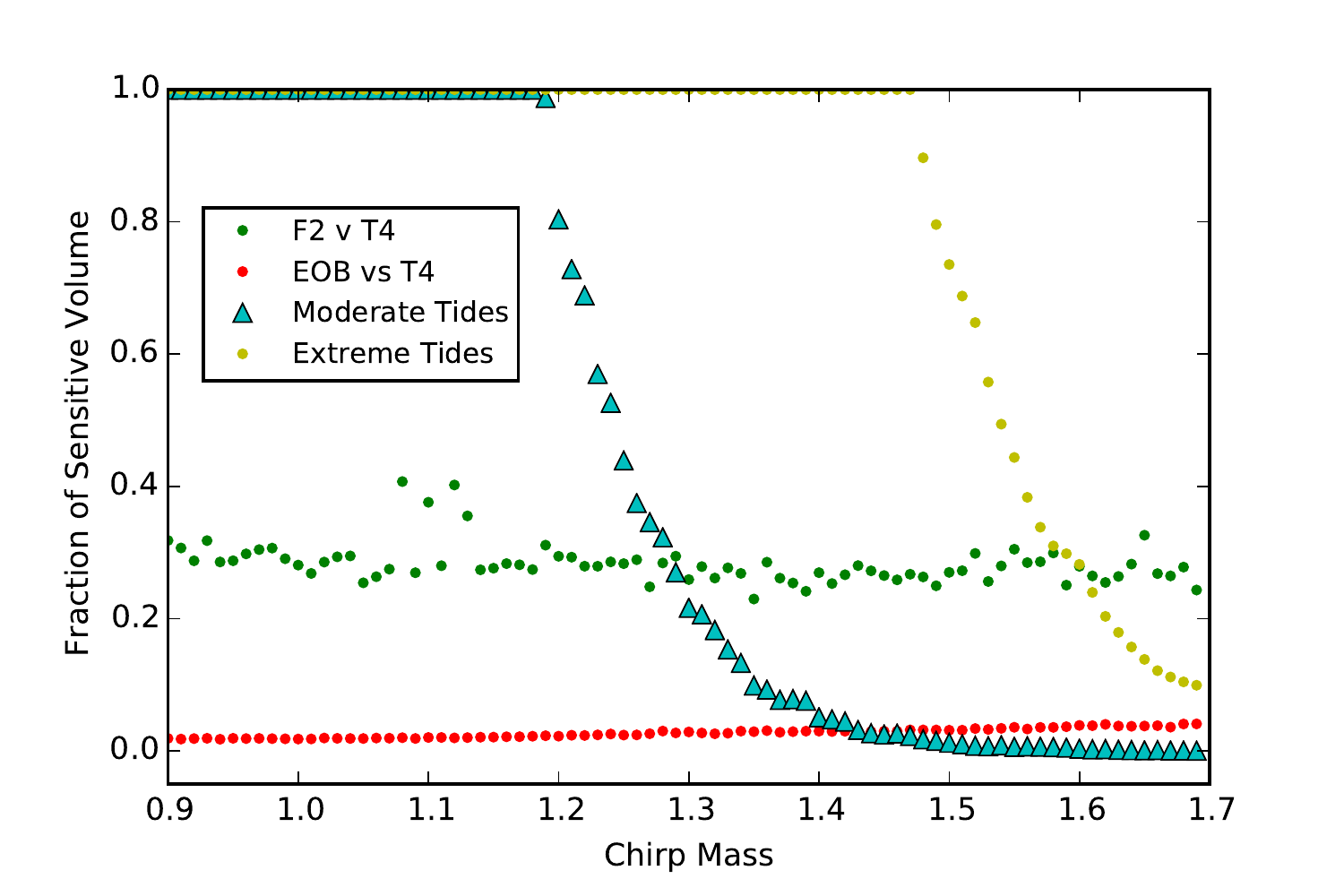}
\caption{The fraction of the detector's sensitive volume where differences between the indicated waveform models would be indistinguishable, as a function of chirp mass. The differences between matter-free approximants are indistinguishable for 70\% or more of the volume, but for low-mass systems, with an EOS-dependent mass cutoff, all detected signals may be affected by tides.}
\label{fig:volume}
\end{figure*}

The maximum distinguishable distances can be compared to the horizon distance to estimate what fraction of detected signals may be affected by a given waveform distance. The horizon distance characterizes a volume containing detectable sources with the detector at the center. Similarly, the distinguishable distances in Figure $\ref{fig:disting}$ characterize a volume of space within which the two systems being compared are  distinguishable. By taking the ratio of these two volumes, the fraction of the sensitive volume of the detectors where two waveforms being compared are distinguishable can be found by 
\begin{eqnarray}
\frac{D^3_{distinguishable}}{D^3_{horizon}}.
\end{eqnarray}
This quantity is plotted as a function of the chirp mass in Figure $\ref{fig:volume}$. Note that when the distinguishable distance goes past the detectable distance, as is the case for the tidal curves in Figure \ref{fig:disting}, this means these two systems are distinguishable up to the horizon distance. This is shown in Figure \ref{fig:volume} by the two curves reaching 1 on the fraction of sensitive volume where the two waveforms would be distinguishable.

We first note that the semi-analytic comparisons, EOB and TaylorF2 to TaylorT4, are below the maximum detectable distance for all masses.  In particular, the EOB vs T4 (red) comparison shows the maximum distinguishable distance for these two approximants ranges from $\approx 50 - 100$ Mpc. This is roughly $\frac{1}{4}$ to $\frac{1}{3}$ of the horizon distance for any given mass combination. When converted to volumes, this indistinguishable region corresponds to 98\% to 96\% of LIGO's total sensitive volume. Therefore, for this study, the difference between EOB and T4 approximants is insignificant. However, the comparison of the two \ac{PN} approximants TaylorT4 and TaylorF2 shows somewhat larger impact, affecting $\sim$30\% of LIGO's sensitive volume--this is in line with previous studies showing that systematic error from choice of \ac{PN} approximants is comparable in size to tidal effects for typical-mass systems. 

The most significant waveform modeling effects that emerge from this study are the leading-order tidal effects. The impact of neglecting tidal effects is shown in Figure $\ref{fig:disting}$ for a moderate (APR4) and extreme (2H) \ac{EOS}. For lower mass systems, these curves were found to be marginally distinguishable well past the horizon distance. In particular, it seems possible for large tidal waveform effects to bias parameter estimation for all detected signals with support in the low-mass region of the \ac{BNS} parameter space. Given the significance of this difference, we also further investigate the tidal versus non-tidal cases in terms of potential search impacts in the following sections.

The final check we make is to evaluate the relative impact of numerical merger effects by introducing hybrid waveforms. These hybrids come from attaching the 33 numerical merger simulation waveforms of \cite{hotokezaka1} to a tidal TaylorT4 waveform with the the same parameters, and comparing that hybrid to a tidal TaylorT4 alone.  The waveform differences measured are thus the matter effects that come from the numerical merger and post-merger. The maximum distinguishable distances for numerical waveforms, marked in black in Figure $\ref{fig:disting}$, are comparable to the range of significance of varying the \ac{PN} approximants; they are comparable to the size of the pure \ac{PN} tidal effects for the moderate APR4 \ac{EOS} and will be important to assess for systematic error in parameter estimation.

\section{Matter impact on Matched Filtering}

We next turn to assess in more detail the effect of matter, using the dominant tidal contributions established in the previous section, on the matched-filtering of signals. This is a method used to find unknown signals in some noise  by comparing it with a known signal, or template. Candidates are identified using the filter output $\left\langle s | h \right\rangle$ (Eq. \ref{eq:prod}) with known signal $s$ and a normalized template $h$. 

\subsection{Match calculations}
We first evaluate the match between two  waveforms $h_1$ and  $h_2$ for comparison by taking the overlap
\begin{eqnarray}
\mathcal{O}(h_1|h_2) = \frac{\langle{h_1}|h_2\rangle{}}{\sqrt[]{\langle{h_1}|h_1\rangle{}\langle{h_2}|h_2\rangle{}}}.
\end{eqnarray}
The match is then defined by maximizing the overlap over the extrinsic arrival time and phase of the signal,

\begin{eqnarray}
\mathcal{M}(h_1|h_2) = \max_{\phi,t} \mathcal{O}(h_1|h_2(\phi,t)).
\end{eqnarray}

This quantity varies between 0 and 1, with 1 indicating that the two waveforms are identical and 0 indicating they are orthogonal. 

\begin{figure}[ht] 
\includegraphics[width=7.0in]{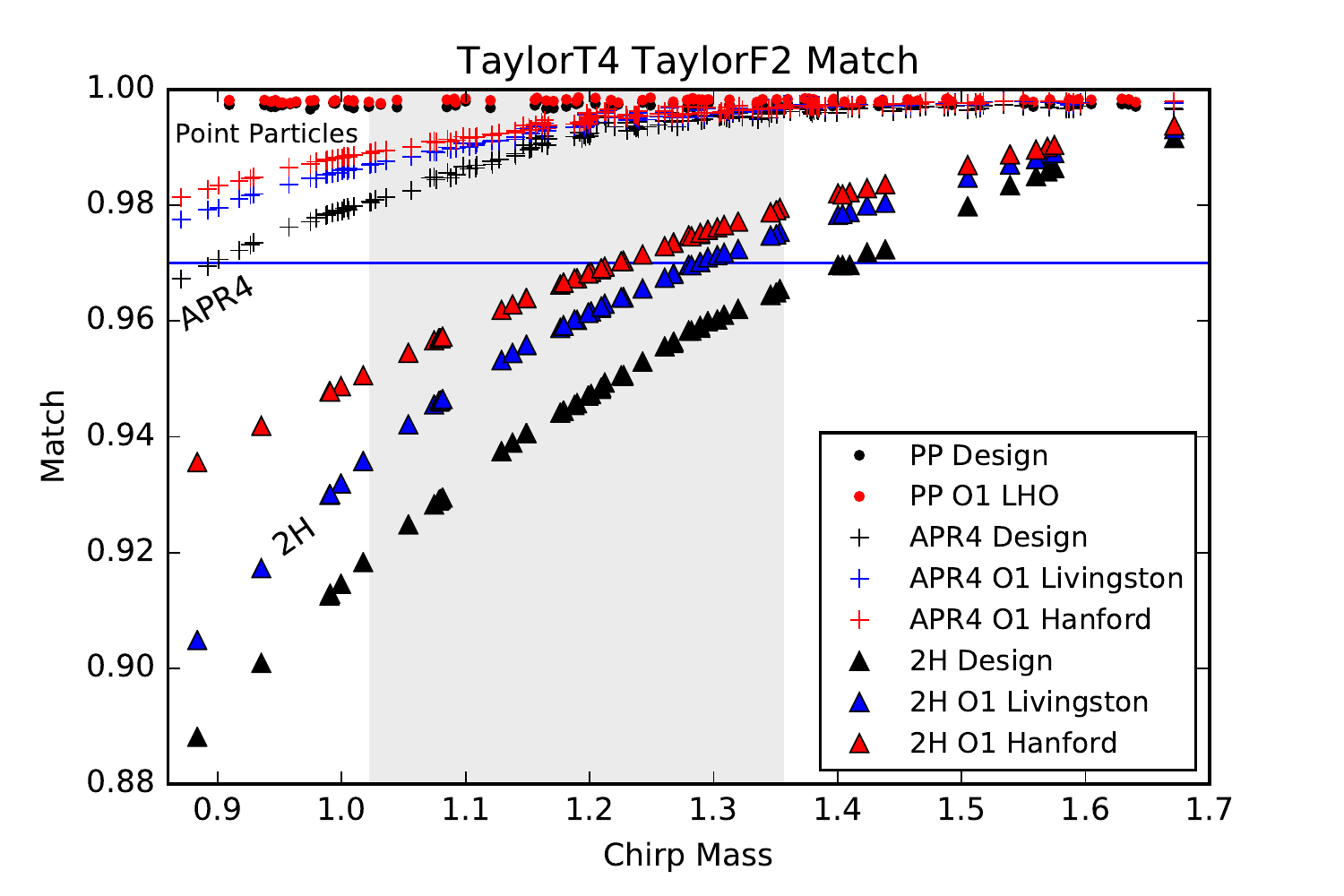}
\caption{Match of Tidal TaylorT4 with TaylorF2 as a function of chirp mass of the system. LIGO's standard 3\% maximum mismatch is illustrated by the blue line. The extreme EOS is below this for a large range of chirp masses, while the moderate EOS is below for the lowest mass systems at design sensitivity. The integral quantities calculated in this figure used a low cutoff frequency of 30 Hz. The gray region represents the chirp mass of previously observed neutron stars \cite{2016ARA&A..54..401O}.}
\label{fig:match}
\end{figure}

We can make a first estimate of how many signals could be lost by recovering signals with a template bank that does not include tidal effects. By taking the match between a tidal TaylorT4 waveform with the existing template bank in a point by point comparison, we can estimate the percentage loss in signals by calculating $1-(1-\mathcal{M})^3$, where $\mathcal{M}$ is the match \cite{2008PhRvD..78l4020L}. A match below the fiducial $0.97$ that sets template-bank spacing by allowing for $\sim10$\% loss in signals indicates a potentially significant effect. A more realistic assessment of search impact will be found with a more computationally intensive template bank study of Section \ref{sec:banksim}.

Figure $\ref{fig:match}$ shows how well the TaylorT4 waveform with tides matches with the corresponding TaylorF2 template with identical mass parameters. 
Neutron-star \ac{EOS} are used to calculate the tidal deformability for the TaylorT4 signals. Results are labelled in Figure $\ref{fig:match}$ as point-particle (no tides), moderate EOS (APR4), and extreme EOS (2H), where the 2H EOS effectively gives an upper bound on the tidal effects seen in realistic EOS.

Figure $\ref{fig:match}$ shows that even in O1 sensitivity, there could have been a significant loss in signals for the lower mass systems if the extreme 2H \ac{EOS} described astrophysical neutron-star matter. Furthermore, as LIGO gets more sensitive at higher frequencies at design sensitivity, the fraction of potentially lost signals increases.  

It is worth noting that the match is not directly linked to the overall sensitivity (i.e. the horizon distance) of a given detector, but depends on the relative sensitivity at different frequencies. This is notable in explaining the differences in match using the O1 Livingston and O1 Hanford noise curves. Hanford was more sensitive than Livingston in the low frequency region below $\sim 100$\,Hz, so this region is more heavily weighted in matches for Hanford than for Livingston. Since tidal effects come in at high frequencies, they had a smaller relative impact with the Hanford sensitivity curve. Both O1 noise curves refer to the average measured sensitivity during Sept $17^{th}$ to Oct $20^{th}$, 2015 \cite{noisecurveref}.

\begin{figure}[htb]
\hspace*{-1cm} 
	\includegraphics[width=7in]{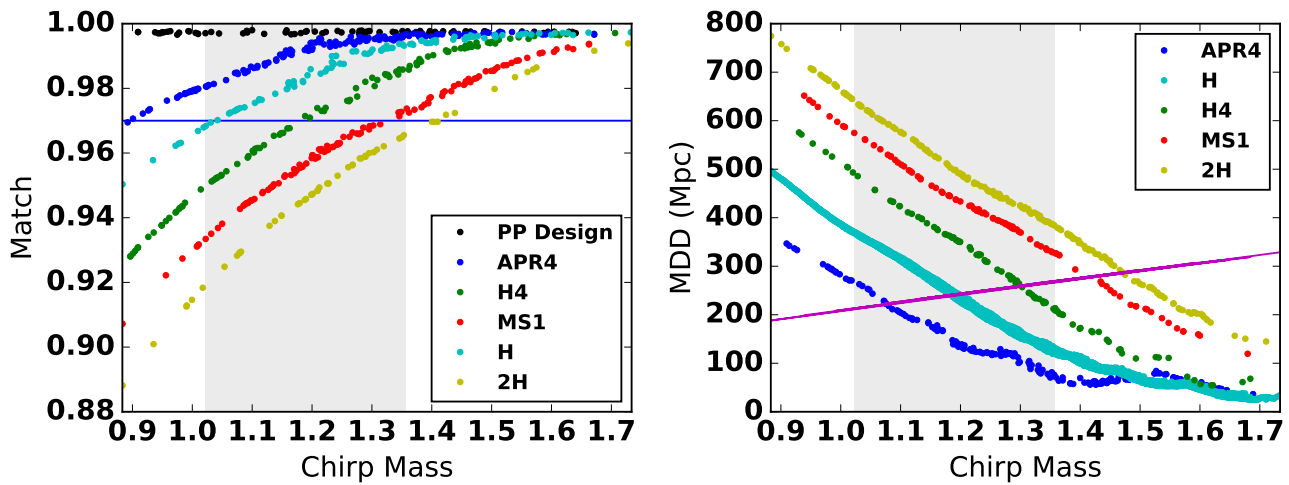}
\caption{Calculation of matches (left) and maximum distinguishable distance (MDD) (right) for the H4, MS1, and H equations of state as a function of chirp mass. All calculations done in this figure are using design sensitivity noise curves.}
\label{fig:disting_match_compare}
\end{figure}

The variation in match and distinguishable distance that lie between the two given Equations of State so far is illustrated in Figure \ref{fig:disting_match_compare} by showing a range of \ac{EOS}. The original curves for the APR4 and 2H equations of state are plotted, with the addition of three others to fill in the approximate range of moderate to extreme tidal effects. In the left subplot of Figure \ref{fig:disting_match_compare} we see the same match calculation as Figure \ref{fig:match} and in the right subplot we see the same distinguishability calculations as Figure \ref{fig:disting}. This  reinforces the sensitivity of our results to the still-unknown neutron-star \ac{EOS}.

\subsection{Template bank recovery}
\label{sec:banksim}
In practice, gravitational waves are not recovered using a template targeted precisely to their mass and other parameters, but by a template bank constructed to discretely sample a range of masses and spins. Here, we use the template bank used by LIGO to search for \ac{BNS} in the first observing run, which contains templates in the \ac{BNS} mass range $[1,2]M_\odot$ with dimensionless spins ranging from $[0,0.05]$ \cite{ratespaper}, and quantitatively estimate how well it would have captured signals with moderate or extreme tidal contributions from matter effects. 
\begin{figure*}[hb] 
\hspace*{-.5cm} 
\includegraphics[width=7.0in]{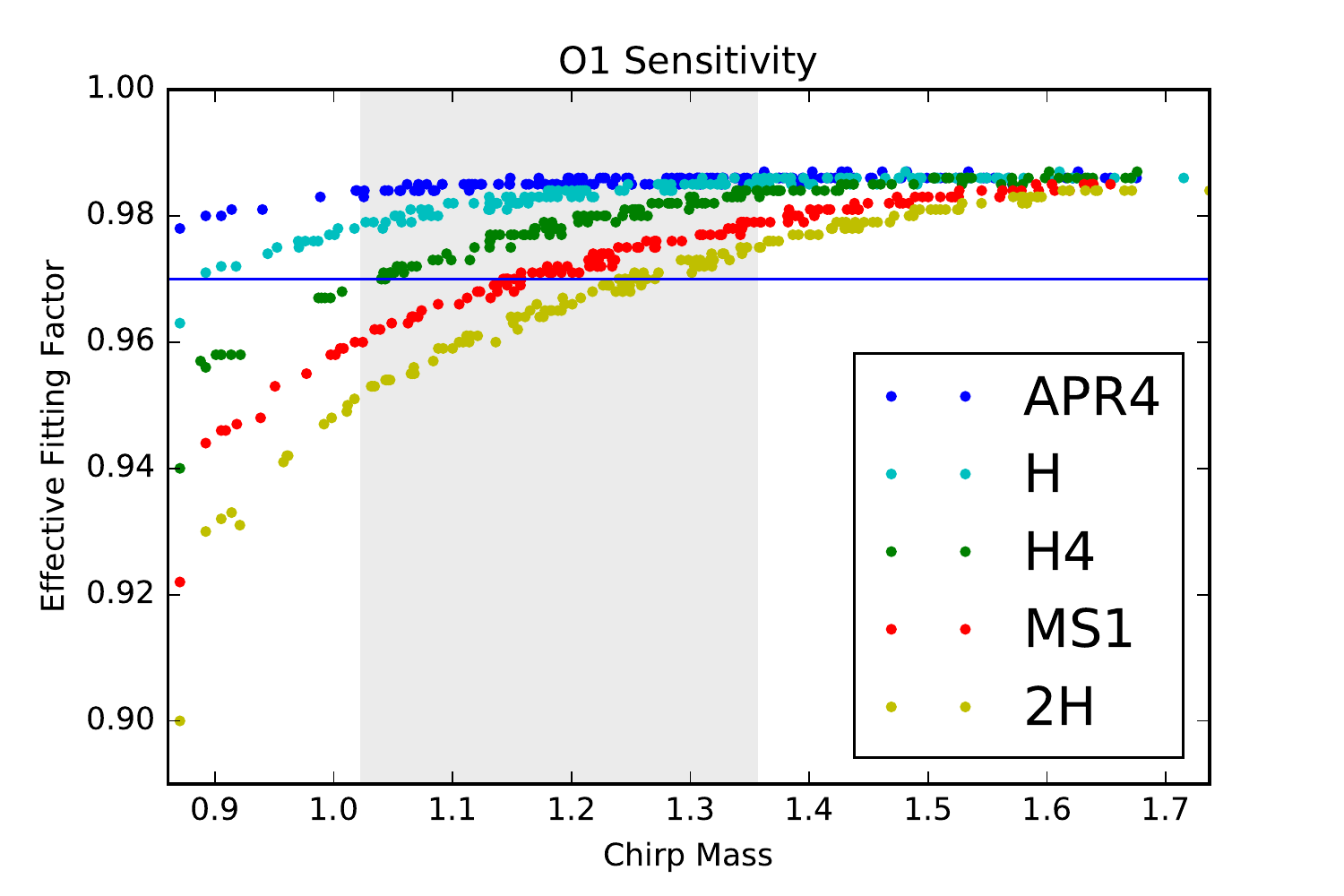}
\caption{Injections of waveforms at fixed masses with matter effects predicted by a range of equations of state. The template bank has been constructed with a minimal fitting factor of 0.97, leading to an averaged ``effective'' fitting factor of roughly 0.985 in the absence of matter effects.
The effective fitting factor for each point injection is plotted as a function of chirp mass for different equations of state. The effective fitting factor was calculated using noise curves from O1\cite{noisecurveref2}.}
\label{fig:eff_chirp_O1}
\end{figure*}
The evaluation of template banks involves injecting simulated signals into noise and determining how well the template bank was able to recover them \cite{2017arXiv170501845D}. 
The sensitivity of this bank to a gravitational waveform $h_s$ with unknown parameters can be characterized by the fitting factor \cite{banksim1,banksim2}
\begin{eqnarray}
FF(h_s) = \max_{h \in \{h_b\}} \mathcal{M}(h_s,h_t).
\end{eqnarray} 

The fitting factor is the match maximized over all templates in the bank. This determines the maximum possible SNR with which a particular waveform can be recovered, but does not necessarily identify the true parameters of the system.


\begin{figure*}[hb] 
\hspace*{-.5cm} 
\includegraphics[width=7.0in]{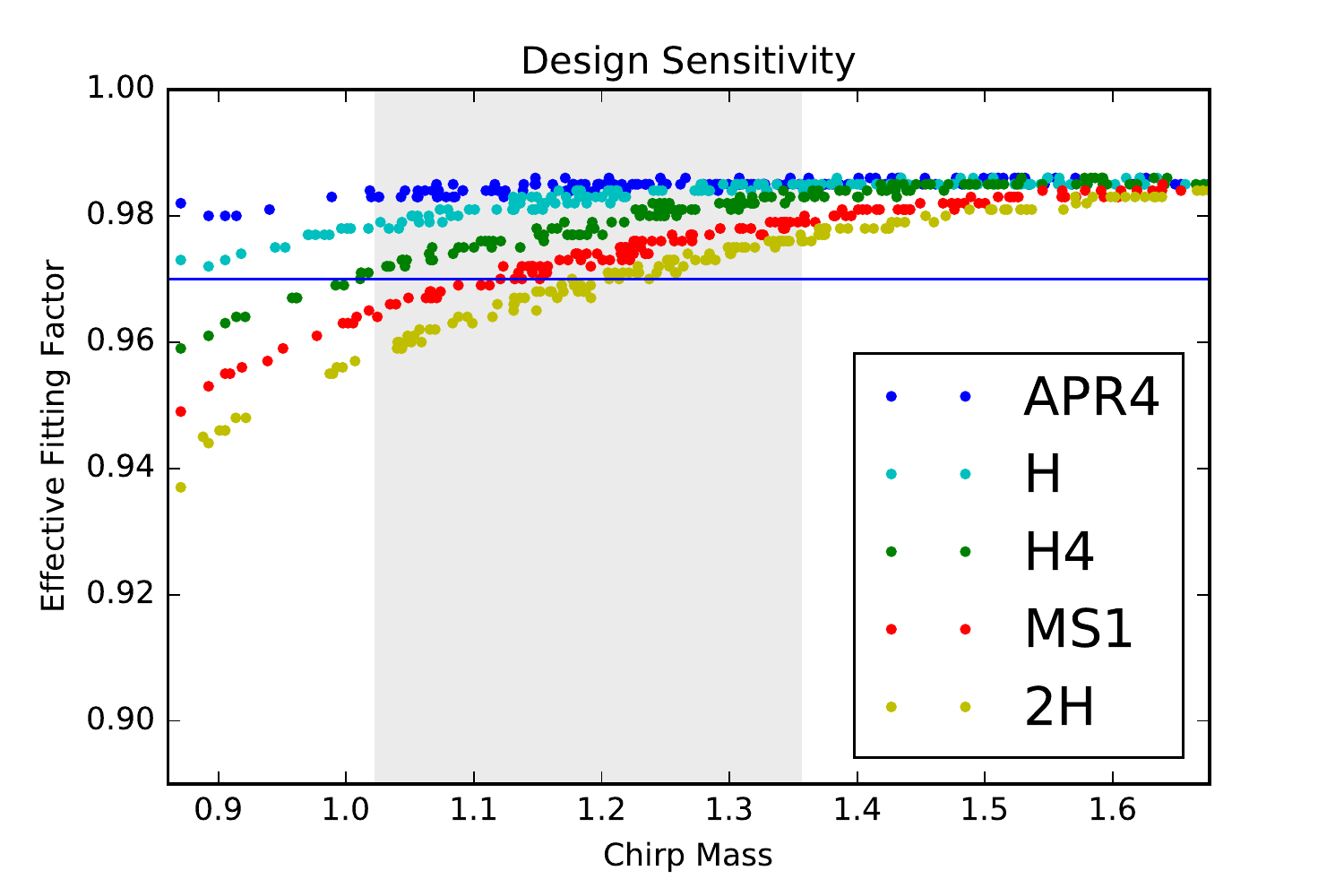}
\caption{Injections of waveforms at fixed masses  with matter effects predicted by a range of equations of state. As in Fig.~\ref{fig:eff_chirp_O1}, the template bank has been constructed with a minimal fitting factor of 0.97, yielding an effective fitting factor at each point of 0.984. The effective fitting factor for each point injection is plotted as a function of chirp mass for different equations of state for design sensitivity noise curves. }
\label{fig:eff_chirp_design}
\end{figure*}

\begin{figure}[htb]
\hspace*{-1cm} 
	\includegraphics[width=3.5in]{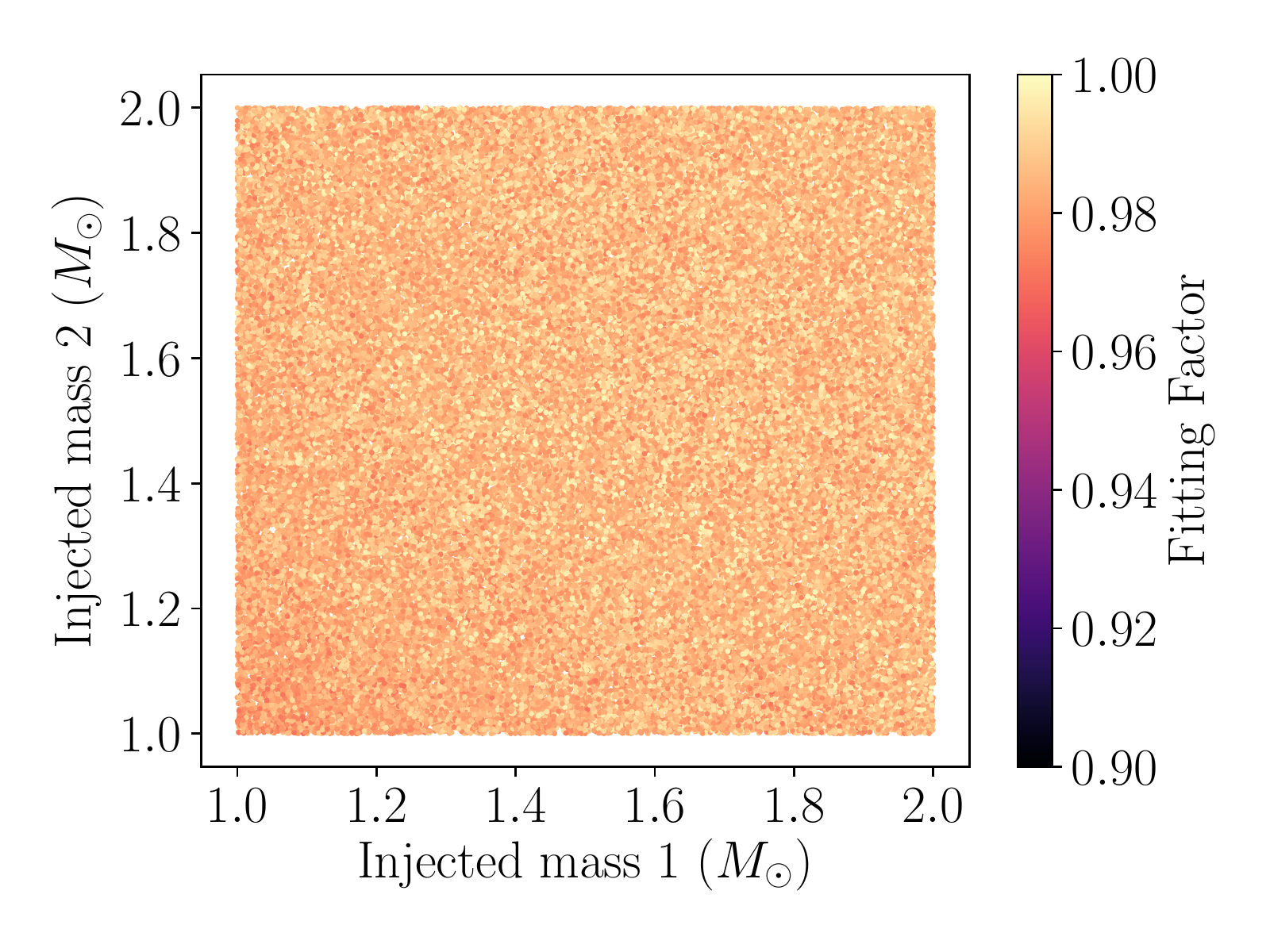}
    \includegraphics[width=3.5in]{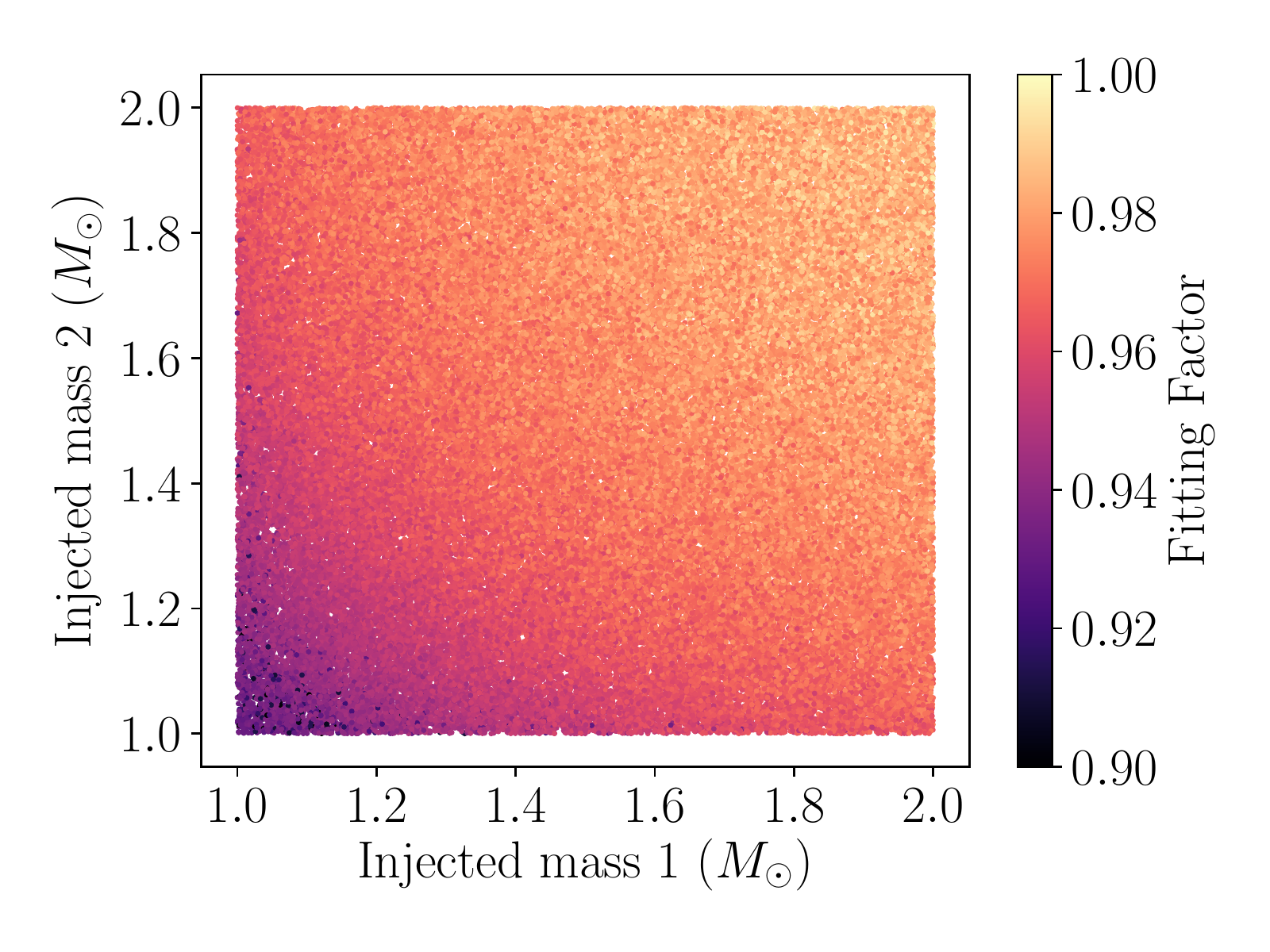}
\caption{Template bank recovery of waveforms with tidal effects at O1 sensitivity using moderate APR4 equation of state (left) and an extreme 2H equation of state (right).}
\label{fig:broadinj}
\end{figure}

In Figure \ref{fig:broadinj} we show how this fitting factor varies as a function of the two component masses. In this Figure we have used an O1 sensitivity curve---an average of the Livingston and Hanford curves used earlier---and chosen 100,000 points with component masses chosen randomly from a uniform distribution between $[1,2]M_\odot$, component dimensionless spin magnitudes are chosen to be aligned and uniformly between $0$ and $0.05$; the orientations and sky location parameters are chosen isotropically. This is shown for both the APR4 and 2H equation-of-states. We see no clear pattern for the moderate EOS (APR4), but a clear mass dependence on the fitting factor when using the 2H equation-of-state.

In Figures \ref{fig:eff_chirp_O1} and \ref{fig:eff_chirp_design} we show an averaged fitting factor as a function of chirp mass. In these plots we have selected 150 unique values of component mass---from a uniform distribution between $[1,2]M_\odot$---and at each point simulated 2000 signals with the same distribution of component dimensionless spins, orientation and sky location parameters used in Figure \ref{fig:broadinj}. We then take the average of these fitting factor values before plotting the data. In Figure \ref{fig:eff_chirp_O1} we use the O1 averaged sensitivity curve and plot results for the five equation-of-states that we consider in this work, and in Figure \ref{fig:eff_chirp_design} we show the same, but instead using the Advanced LIGO design sensitivity curve.

The rate of astrophysical events $R$ can be related to the observed rate of BNS events $\Lambda$ by
\begin{eqnarray}
\Lambda = R \langle{VT} \rangle,
\end{eqnarray}
where the quantity $\langle{VT} \rangle$ is the volume of space-time that the detectors are sensitive to. Abbott et al \cite{ratespaper} note that if the effect of tides is extreme, the true sensitive volume $\langle{VT} \rangle$ will be smaller than calculated, roughly by a factor of the fitting factor cubed for injections of the systems considered. We find here that this is limited by the matter effects for the lowest-mass 1.0,1.0$M_\odot$ systems if typical radii are between 12 and 13 km, as in the H EOS. If neutron-star radii are larger, around the 14 km of the 2H equation of state, all neutron-star chirp masses observed in galactic binaries will have noticeable matter effects on the template bank fitting factors, with particularly large impact (fitting factor down to 0.9) at low mass or with even larger radii like the 16 km of the 2H equation of state. This would translate into a low-mass rate constraint that is only $0.9^3\simeq0.73$ as strong as estimated using point-particle templates, an error of up to $\sim 27\%$. Strong tidal effects would have a significant impact on estimated $\langle VT \rangle$, however, note that the upper limits on rates from the first observing run \cite{ratespaper} have competing uncertainties of $18\%$ from calibration uncertainties and $40\%$ due to choices of prior. 

\section{Conclusions}
Using approximant and hybrid waveforms we have shown that tidal effects are important for BNS searches with advanced LIGO, and are more important than inter-approximant differences or numerical effects. We show also that tidal effects are particularly significant for low-mass systems and should be considered during parameter estimation. We estimate the impact on searches by calculating the recovery of signals using realistic template banks and find that for extreme EOSs, signal loss from tidal effects is significant in O1 and will continue to matter at Advanced LIGO's design sensitivity. Even for moderate EOS, some signal loss is expected for the lowest mass systems. 

\section{Acknowledgements}
This material is based upon work supported by the National Science Foundation under Grant No. 1307545. Computational resources were provided by the ORCA cluster at California
State University, Fullerton (CSUF), supported by
CSUF, NSF grant No. PHY-142987, and the Research
Corporation for Science Advancement. Results were generated using the PyCBC software package\cite{Canton:2014ena,Usman:2015kfa,pycbc-software}. The authors thank Kenta Hotokezaka, Koutarou Kyutoku, and Masaru Shibata for kindly sharing numerical waveform data.
\clearpage
\appendix
\section{BNS Simulations} 
\begin{figure*}[ht] 
\hspace*{-.5cm} 
\includegraphics[width=7.0in]{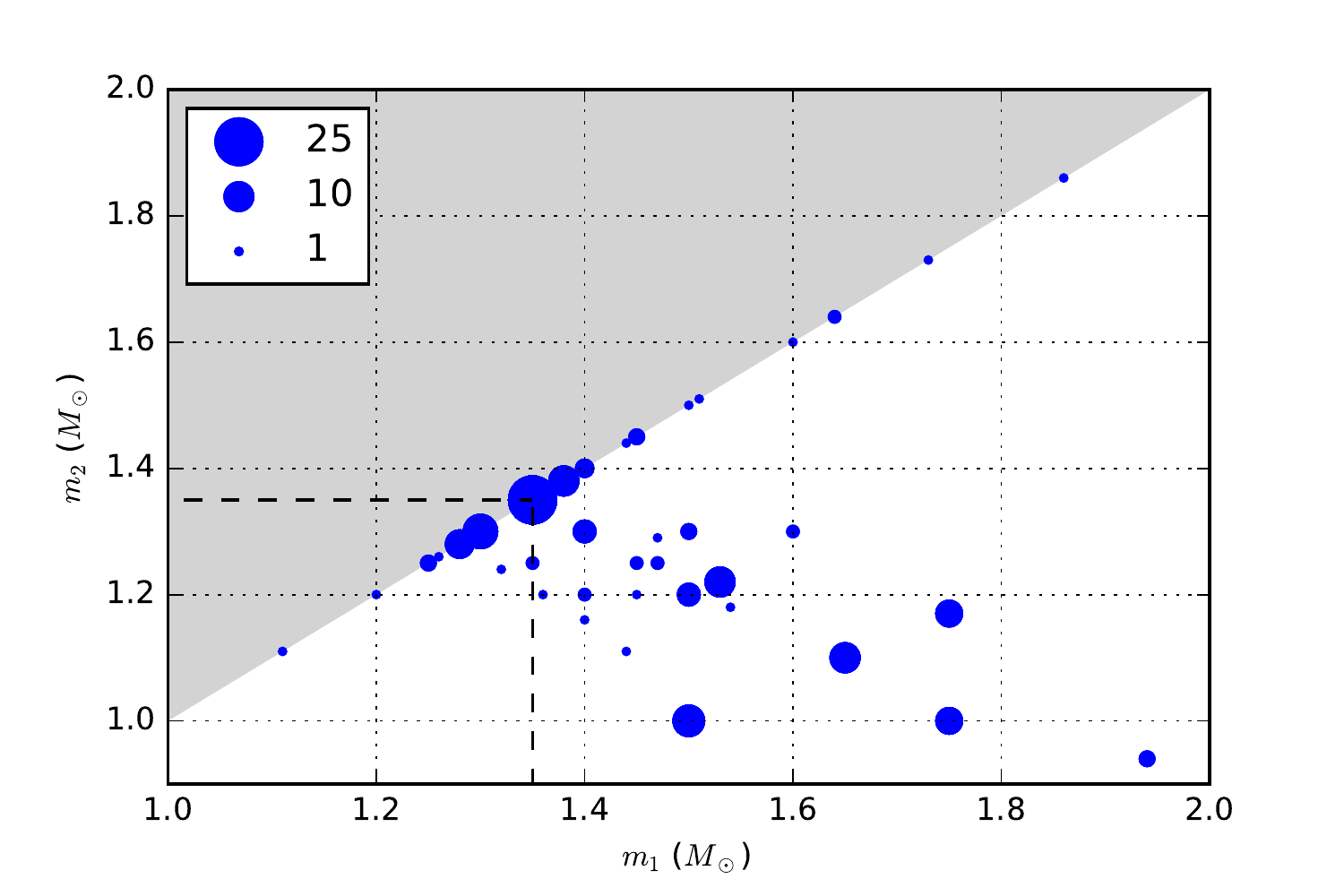}
\caption{A collection of BNS simulations organized by component mass created by taking data from Refs \cite{bns_sims1,bns_sims2,bns_sims3,bns_sims4,bns_sims5,bns_sims6,bns_sims7,bns_sims8,bns_sims9,bns_sims10,bns_sims11,bns_sims12,bns_sims13}. The size of the points indicate how many simulations are known to have that mass combination, with a mode of 1.35, 1.35 $M_\odot$ (occurring 25 times). The data was organized such that $m_1 \geq m_2$ regardless of how it was depicted in the respective work. Note that other simulation parameters such as the different equations of states of the stars (10+), initial gravitational-wave frequency (370-700+ Hz), duration of simulation (ranging from the order of milliseconds to seconds), or any other relevant parameters are not shown.}
\label{fig:bns-sims}
\end{figure*}

\section{Total mass versus chirp mass}
\label{sec:chirpvstotal} We plot characteristic distances as a function of chirp mass. It turns out that the size of the tidal effect is primarily determined as a function of chirp mass, leading to the line-like plots of Figure {\ref{fig:disting}; in contrast, the tidal distinguishability comparisons have a distinct thickness to them when plotted as a function of total mass as seen in Figure $\ref{fig:colormap}$. At a total mass of 3 $M_\odot$, where the potential mass ratio is largest, the distance differs by about 100 Mpc between the cases of 1.5,1.5 $M_\odot$ and 1.0,2.0 $M_\odot$. This means that higher mass ratio systems have leading-order tidal effects that are distinguishable out to a farther distance.
\begin{figure}[htb]
\includegraphics[width=7.0in]{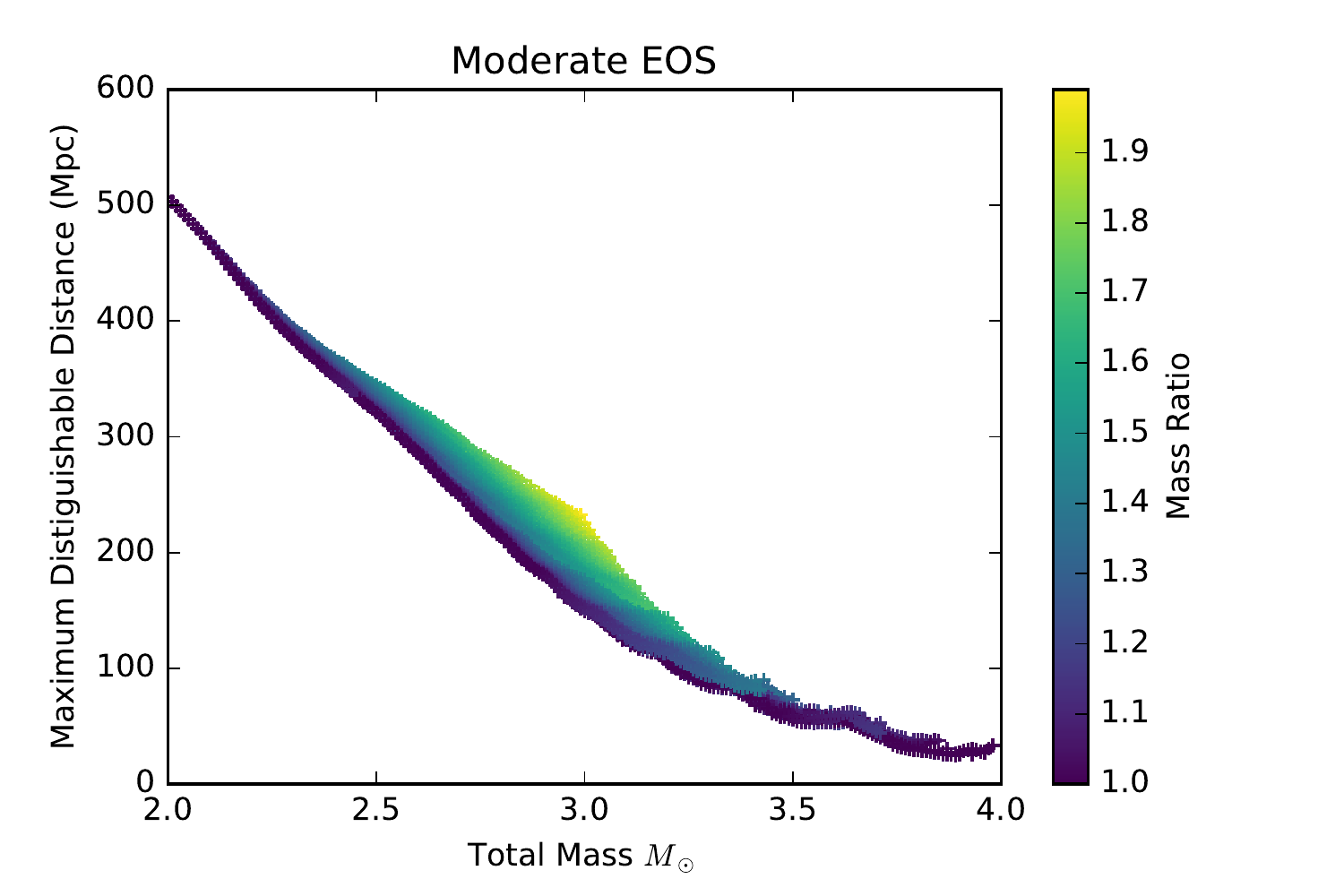}
\caption{The tidal TaylorT4 vs TaylorT4 distinguishability calculation weighted by mass ratio.}
\label{fig:colormap}
\end{figure}

\section*{\refname}
\bibliographystyle{iopart_num}
\bibliography{References}

\providecommand{\newblock}{}
\begin{thebibliography}{10}
\expandafter\ifx\csname url\endcsname\relax
  \def\url#1{{\tt #1}}\fi
\expandafter\ifx\csname urlprefix\endcsname\relax\def\urlprefix{URL }\fi
\providecommand{\eprint}[2][]{\url{#2}}

\bibitem{2016PhRvL.116f1102A}
{Abbott} B~P, {Abbott} R, {Abbott} T~D, {Abernathy} M~R, {Acernese} F, {Ackley}
  K, {Adams} C, {Adams} T, {Addesso} P, {Adhikari} R~X and et~al 2016  {\bf
  116} 061102 (\textit{Preprint} \eprint{1602.03837})

\bibitem{2016PhRvL.116x1103A}
{Abbott} B~P, {Abbott} R, {Abbott} T~D, {Abernathy} M~R, {Acernese} F, {Ackley}
  K, {Adams} C, {Adams} T, {Addesso} P, {Adhikari} R~X and et~al 2016 {\em
  Physical Review Letters\/} {\bf 116} 241103 (\textit{Preprint}
  \eprint{1606.04855})

\bibitem{PhysRevLett.118.221101}
Abbott B~P, Abbott R, Abbott T~D, Acernese F, Ackley K, Adams C, Adams T,
  Addesso P, Adhikari R~X, Adya V~B and et~al (LIGO Scientific and Virgo
  Collaboration) 2017 {\em Phys. Rev. Lett.\/} {\bf 118}(22) 221101
  \urlprefix\url{https://link.aps.org/doi/10.1103/PhysRevLett.118.221101}

\bibitem{TheLIGOScientific:2014jea}
Aasi J {\em et~al.\/} (LIGO Scientific) 2015 {\em Class. Quant. Grav.\/} {\bf
  32} 074001 (\textit{Preprint} \eprint{1411.4547})

\bibitem{TheVirgo:2014hva}
Acernese F {\em et~al.\/} (VIRGO) 2015 {\em Class. Quant. Grav.\/} {\bf 32}
  024001 (\textit{Preprint} \eprint{1408.3978})

\bibitem{PhysRevD.88.043007}
Aso Y, Michimura Y, Somiya K, Ando M, Miyakawa O, Sekiguchi T, Tatsumi D and
  Yamamoto H (The KAGRA Collaboration) 2013 {\em Phys. Rev. D\/} {\bf 88}(4)
  043007 \urlprefix\url{https://link.aps.org/doi/10.1103/PhysRevD.88.043007}

\bibitem{2016PhRvX...6d1015A}
{Abbott} B~P, {Abbott} R, {Abbott} T~D, {Abernathy} M~R, {Acernese} F, {Ackley}
  K, {Adams} C, {Adams} T, {Addesso} P, {Adhikari} R~X and et~al 2016 {\em
  Physical Review X\/} {\bf 6} 041015 (\textit{Preprint} \eprint{1606.04856})

\bibitem{ratespaper}
{Abbott} B~P, {Abbott} R, {Abbott} T~D, {Abernathy} M~R, {Acernese} F, {Ackley}
  K, {Adams} C, {Adams} T, {Addesso} P, {Adhikari} R~X and et~al 2016  {\bf
  832} L21 (\textit{Preprint} \eprint{1607.07456})

\bibitem{2016LRR....19....1A}
{Abbott} B~P, {Abbott} R, {Abbott} T~D, {Abernathy} M~R, {Acernese} F, {Ackley}
  K, {Adams} C, {Adams} T, {Addesso} P, {Adhikari} R~X and et~al 2016 {\em
  Living Reviews in Relativity\/} {\bf 19} 1 (\textit{Preprint}
  \eprint{1304.0670})

\bibitem{2017arXiv170501845D}
{Dal Canton} T and {Harry} I~W 2017 {\em ArXiv e-prints\/} (\textit{Preprint}
  \eprint{1705.01845})

\bibitem{2017arXiv170404628T}
{The LIGO Scientific Collaboration}, {the Virgo Collaboration}, {Abbott} B~P,
  {Abbott} R, {Abbott} T~D, {Acernese} F, {Ackley} K, {Adams} C, {Adams} T,
  {Addesso} P and et~al 2017 {\em ArXiv e-prints\/} (\textit{Preprint}
  \eprint{1704.04628})

\bibitem{1992ApJ...398..234K}
{Kochanek} C~S 1992 {\em \apj\/} {\bf 398} 234--247

\bibitem{1992ApJ...400..175B}
{Bildsten} L and {Cutler} C 1992 {\em \apj\/} {\bf 400} 175--180

\bibitem{1996PhRvD..54.3958L}
{Lai} D and {Wiseman} A~G 1996 {\em \prd\/} {\bf 54} 3958--3964
  (\textit{Preprint} \eprint{gr-qc/9609014})

\bibitem{1993PhRvL..70.2984C}
{Cutler} C, {Apostolatos} T~A, {Bildsten} L, {Finn} L~S, {Flanagan} E~E,
  {Kennefick} D, {Markovic} D~M, {Ori} A, {Poisson} E and {Sussman} G~J 1993
  {\em Physical Review Letters\/} {\bf 70} 2984--2987 (\textit{Preprint}
  \eprint{astro-ph/9208005})

\bibitem{2012PhRvD..85l2006A}
{Allen} B, {Anderson} W~G, {Brady} P~R, {Brown} D~A and {Creighton} J~D~E 2012
  {\em \prd\/} {\bf 85} 122006 (\textit{Preprint} \eprint{gr-qc/0509116})

\bibitem{2002PhRvD..66f4013B}
{Berti} E, {Pons} J~A, {Miniutti} G, {Gualtieri} L and {Ferrari} V 2002 {\em
  \prd\/} {\bf 66} 064013 (\textit{Preprint} \eprint{gr-qc/0208011})

\bibitem{2008PhRvD..77b1502F}
{Flanagan} {\'E}~{\'E} and {Hinderer} T 2008 {\em \prd\/} {\bf 77} 021502
  (\textit{Preprint} \eprint{0709.1915})

\bibitem{2016ARA&A..54..401O}
{{\"O}zel} F and {Freire} P 2016 {\em \araa\/} {\bf 54} 401--440
  (\textit{Preprint} \eprint{1603.02698})

\bibitem{2015ApJ...812..143M}
{Martinez} J~G, {Stovall} K, {Freire} P~C~C, {Deneva} J~S, {Jenet} F~A,
  {McLaughlin} M~A, {Bagchi} M, {Bates} S~D and {Ridolfi} A 2015 {\em \apj\/}
  {\bf 812} 143 (\textit{Preprint} \eprint{1509.08805})

\bibitem{LALref}
{LIGO Scientific Collaboration} 2017 Lalsuite install instructions
  \urlprefix\url{https://wiki.ligo.org/DASWG/LALSuiteInstall}

\bibitem{T4&F2}
{Buonanno} A, {Iyer} B~R, {Ochsner} E, {Pan} Y and {Sathyaprakash} B~S 2009
  {\em PRD\/} {\bf 80} 084043 (\textit{Preprint} \eprint{0907.0700})

\bibitem{2016PhRvD..93d4064B}
{Barkett} K, {Scheel} M~A, {Haas} R, {Ott} C~D, {Bernuzzi} S, {Brown} D~A,
  {Szil{\'a}gyi} B, {Kaplan} J~D, {Lippuner} J, {Muhlberger} C~D, {Foucart} F
  and {Duez} M~D 2016 {\em \prd\/} {\bf 93} 044064 (\textit{Preprint}
  \eprint{1509.05782})

\bibitem{EOB}
{Pan} Y, {Buonanno} A, {Boyle} M, {Buchman} L~T, {Kidder} L~E, {Pfeiffer} H~P
  and {Scheel} M~A 2011 {\em PRD\/} {\bf 84} 124052 (\textit{Preprint}
  \eprint{1106.1021})

\bibitem{hotokezaka1}
{Hotokezaka} K, {Kyutoku} K and {Shibata} M 2013  {\bf 87} 044001
  (\textit{Preprint} \eprint{1301.3555})

\bibitem{wadespaper}
{Wade} L, {Creighton} J~D~E, {Ochsner} E, {Lackey} B~D, {Farr} B~F,
  {Littenberg} T~B and {Raymond} V 2014 {\em \prd\/} {\bf 89} 103012
  (\textit{Preprint} \eprint{1402.5156})

\bibitem{readspaper}
{Read} J~S, {Baiotti} L, {Creighton} J~D~E, {Friedman} J~L, {Giacomazzo} B,
  {Kyutoku} K, {Markakis} C, {Rezzolla} L, {Shibata} M and {Taniguchi} K 2013
  {\bf 88} 044042 (\textit{Preprint} \eprint{1306.4065})

\bibitem{2004PhRvD..69j4017P}
{Pan} Y, {Buonanno} A, {Chen} Y and {Vallisneri} M 2004 {\em PRD\/} {\bf 69}
  104017 (\textit{Preprint} \eprint{gr-qc/0310034})

\bibitem{2009PhRvD..79l4032R}
{Read} J~S, {Lackey} B~D, {Owen} B~J and {Friedman} J~L 2009  {\bf 79} 124032
  (\textit{Preprint} \eprint{0812.2163})

\bibitem{hotokezaka2}
{Hotokezaka} K, {Kyutoku} K, {Okawa} H, {Shibata} M and {Kiuchi} K 2011  {\bf
  83} 124008 (\textit{Preprint} \eprint{1105.4370})

\bibitem{2008PhRvD..78l4020L}
{Lindblom} L, {Owen} B~J and {Brown} D~A 2008  {\bf 78} 124020
  (\textit{Preprint} \eprint{0809.3844})

\bibitem{noisecurveref}
{Abbott} B~P, {Abbott} R, {Abbott} T~D, {Abernathy} M~R, {Acernese} F, {Ackley}
  K, {Adamo} M, {Adams} C, {Adams} T, {Addesso} P and et~al 2016 {\em Classical
  and Quantum Gravity\/} {\bf 33} 134001 data can be found on the DCC from
  document number LIGO-T1600030-v2 (\textit{Preprint} \eprint{1602.03844})

\bibitem{noisecurveref2}
{Barsotti} L and P F  Tech. Rep. T1200307, The LIGO Scientific Collaboration
  and the Virgo Collaboration (2017), https://dcc.ligo.org/LIGO-T1200307/
  public.

\bibitem{banksim1}
{Brown} D~A, {Harry} I, {Lundgren} A and {Nitz} A~H 2012 {\em \prd\/} {\bf 86}
  084017 (\textit{Preprint} \eprint{1207.6406})

\bibitem{banksim2}
{Harry} I~W, {Nitz} A~H, {Brown} D~A, {Lundgren} A~P, {Ochsner} E and {Keppel}
  D 2014 {\em \prd\/} {\bf 89} 024010 (\textit{Preprint} \eprint{1307.3562})

\bibitem{Canton:2014ena}
Dal~Canton T {\em et~al.\/} 2014 {\em Phys. Rev.\/} {\bf D90} 082004
  (\textit{Preprint} \eprint{1405.6731})

\bibitem{Usman:2015kfa}
Usman S~A {\em et~al.\/} 2016 {\em Class. Quant. Grav.\/} {\bf 33} 215004
  (\textit{Preprint} \eprint{1508.02357})

\bibitem{pycbc-software}
Nitz A, Harry I, Biwer C~M, Brown D, Willis J, Canton T~D, Pekowsky L, Dent T,
  Williamson A~R, Capano C, De S, Machenschalk B, Kumar P, Cabero M, Massinger
  T, Lenon A, Fairhurst S, Reyes S, Nielsen A, shasvath, Pannarale F, Singer L,
  Macleod D, Babak S, Gabbard H, Sugar C, Zertuche L~M, Khan S, couvares and
  Bockelman B 2017 ligo-cbc/pycbc: O2 production release 12
  \urlprefix\url{https://doi.org/10.5281/zenodo.809404}

\bibitem{bns_sims1}
{Lehner} L, {Liebling} S~L, {Palenzuela} C, {Caballero} O~L, {O'Connor} E,
  {Anderson} M and {Neilsen} D 2016 {\em Classical and Quantum Gravity\/} {\bf
  33} 184002 (\textit{Preprint} \eprint{1603.00501})

\bibitem{bns_sims2}
{Feo} A, {De Pietri} R, {Maione} F and {L{\"o}ffler} F 2017 {\em Classical and
  Quantum Gravity\/} {\bf 34} 034001 (\textit{Preprint} \eprint{1608.02810})

\bibitem{bns_sims3}
{De Pietri} R, {Feo} A, {Maione} F and {L{\"o}ffler} F 2016 {\em \prd\/} {\bf
  93} 064047 (\textit{Preprint} \eprint{1509.08804})

\bibitem{bns_sims4}
{Maione} F, {De Pietri} R, {Feo} A and {L{\"o}ffler} F 2016 {\em Classical and
  Quantum Gravity\/} {\bf 33} 175009 (\textit{Preprint} \eprint{1605.03424})

\bibitem{bns_sims5}
{Dietrich} T and {Hinderer} T 2017 {\em ArXiv e-prints\/} (\textit{Preprint}
  \eprint{1702.02053})

\bibitem{bns_sims6}
{Dietrich} T, {Bernuzzi} S and {Tichy} W 2017 {\em ArXiv e-prints\/}
  (\textit{Preprint} \eprint{1706.02969})

\bibitem{bns_sims7}
{Bernuzzi} S, {Nagar} A, {Dietrich} T and {Damour} T 2015 {\em Physical Review
  Letters\/} {\bf 114} 161103 (\textit{Preprint} \eprint{1412.4553})

\bibitem{bns_sims8}
{Hotokezaka} K, {Kyutoku} K, {Sekiguchi} Y~i and {Shibata} M 2016 {\em \prd\/}
  {\bf 93} 064082 (\textit{Preprint} \eprint{1603.01286})

\bibitem{bns_sims9}
{Hinderer} T, {Taracchini} A, {Foucart} F, {Buonanno} A, {Steinhoff} J, {Duez}
  M, {Kidder} L~E, {Pfeiffer} H~P, {Scheel} M~A, {Szilagyi} B, {Hotokezaka} K,
  {Kyutoku} K, {Shibata} M and {Carpenter} C~W 2016 {\em Physical Review
  Letters\/} {\bf 116} 181101 (\textit{Preprint} \eprint{1602.00599})

\bibitem{bns_sims10}
{Hotokezaka} K, {Kyutoku} K, {Okawa} H and {Shibata} M 2015 {\em \prd\/} {\bf
  91} 064060 (\textit{Preprint} \eprint{1502.03457})

\bibitem{bns_sims11}
{Hotokezaka} K, {Kiuchi} K, {Kyutoku} K, {Muranushi} T, {Sekiguchi} Y~i,
  {Shibata} M and {Taniguchi} K 2013 {\em \prd\/} {\bf 88} 044026
  (\textit{Preprint} \eprint{1307.5888})

\bibitem{bns_sims12}
{Haas} R, {Ott} C~D, {Szilagyi} B, {Kaplan} J~D, {Lippuner} J, {Scheel} M~A,
  {Barkett} K, {Muhlberger} C~D, {Dietrich} T, {Duez} M~D, {Foucart} F,
  {Pfeiffer} H~P, {Kidder} L~E and {Teukolsky} S~A 2016 {\em \prd\/} {\bf 93}
  124062 (\textit{Preprint} \eprint{1604.00782})

\bibitem{bns_sims13}
Dietrich T, Ujevic M, Tichy W, Bernuzzi S and Br\"ugmann B 2017 {\em Phys. Rev.
  D\/} {\bf 95}(2) 024029
  \urlprefix\url{https://link.aps.org/doi/10.1103/PhysRevD.95.024029}

\end{thebibliography}
\end{document}